# The Lorentz force law and its connections to hidden momentum, the Einstein-Laub force, and the Aharonov-Casher effect

Masud Mansuripur

College of Optical Sciences, The University of Arizona, Tucson, Arizona 85721

**Abstract**. The Lorentz force of classical electrodynamics, when applied to magnetic materials, gives rise to hidden energy and hidden momentum. Removing the contributions of hidden entities from the Poynting vector, from the electromagnetic momentum density, and from the Lorentz force and torque densities simplifies the equations of the classical theory. In particular, the reduced expression of the electromagnetic force-density becomes very similar (but not identical) to the Einstein-Laub expression for the force exerted by electric and magnetic fields on a distribution of charge, current, polarization and magnetization. Examples reveal the similarities and differences among various equations that describe the force and torque exerted by electromagnetic fields on material media. An important example of the simplifications afforded by the Einstein-Laub formula is provided by a magnetic dipole moving in a static electric field and exhibiting the Aharonov-Casher effect.

**1. Introduction**. The classical theory of electrodynamics is based on Maxwell's equations and the Lorentz force law [1-4]. In their microscopic version, Maxwell's equations relate the electromagnetic (EM) fields, $E(r,t)$ and $B(r,t)$, to the spatio-temporal distribution of electric charge and current densities, $\rho(r,t)$ and $J(r,t)$. In any closed system consisting of an arbitrary distribution of charge and current, Maxwell's equations uniquely determine the field distributions provided that the sources, $\rho$ and $J$, are fully specified in advance. The EM force-density $F_1(r,t)$ and torque-density $T_1(r,t)$ that the $E$ and $B$ fields exert on material media, the seat of $\rho$ and $J$, are then given by the Lorentz law, as follows:

$$F_1(r,t) = \rho(r,t)E(r,t) + J(r,t) \times B(r,t), \tag{1}$$

$$T_1(r,t) = r \times F_1(r,t). \tag{2}$$

(The subscript 1 here distinguishes the Lorentz force and torque from alternative expressions to be introduced later.) The force and torque densities in the above equations are purely electromagnetic. In general, there are other forces and torques that the various parts of the material media exert on each other. The spatio-temporal evolution of $\rho$ and $J$ is thus determined by the action of *total* force and *total* torque on charge carriers. However, since our analysis is based on the assumption that $\rho(r,t)$ and $J(r,t)$ are fully specified *in advance*, we need not be concerned with the internal interactions among the material parts. This is not to ignore the significance of such internal interactions in determining, for instance, the constitutive relations, or the mechanical response of the host media to the EM fields; rather, by hiding away the internal processes and mechanisms that give rise to the specific forms of $\rho(r,t)$ and $J(r,t)$, we manage to focus our attention exclusively on the EM subsystem. The Lorentz force and torque expressions thus specify the rates of exchange of linear and angular momenta between the EM subsystem and the host media. In other words, in the absence of information about interactions that take place within the media, Eq.(1) alone cannot tell the material subsystem how to react to the $E$ and $B$ fields; however, it does inform as to how much momentum per unit volume per unit time the EM fields give (or take away from) the material subsystem. Similarly, the Lorentz torque-density of Eq.(2) is a statement about the time-rate-of-exchange of angular momentum between the fields and the material media.

Aside from Maxwell's equations and the Lorentz force law, the classical theory contains a few other ingredients, such as the Poynting vector $S_1(r,t) = \mu_o^{-1} E \times B$, and the linear and angular momentum densities $p_{EM\_1}(r,t) = \varepsilon_o E \times B$ and $L_{EM\_1}(r,t) = r \times p_{EM\_1}$. (Here $\varepsilon_o$ and $\mu_o$ are the permittivity and permeability of free space.) The Poynting theorem, derived from Maxwell's equations in conjunction with the Poynting postulate that $S_1(r,t)$ represents the rate of flow of EM energy per unit area per unit time, identifies $\mathcal{E}_1(r,t) = \tfrac{1}{2}\varepsilon_o E \cdot E + \tfrac{1}{2}\mu_o^{-1} B \cdot B$ as the EM energy-density, and $E(r,t) \cdot J(r,t)$ as the rate of exchange of energy with the material subsystem [1]. It is also important to know that $(J, c\rho)$ is a 4-vector, while $E$ and $B$ together form a 2$^{nd}$ rank tensor, both of which can be readily transformed between inertial frames in accordance with the Lorentz transformation rules [1-3]. The scalar and vector potentials $\psi(r,t)$ and $A(r,t)$, to which $E$ and $B$ are related through the identities $E(r,t) = -\nabla\psi - \partial A/\partial t$ and $B(r,t) = \nabla \times A$, also form a 4-vector, $(cA, \psi)$.

Suppose now that the EM sources include polarization $P(r,t)$ and magnetization $M(r,t)$, in addition to the aforementioned $\rho(r,t)$ and $J(r,t)$. Once again, we assume that the spatio-temporal distributions of $\rho$, $J$, $P$ and $M$ are fully specified in advance, so that there is no need for any constitutive relations to enable us to obtain the source distributions from a knowledge of the EM fields. Similarly, we do not need to know the internal forces and torques that the various parts of the media exert upon each other, which would be needed if one were to calculate the spatio-temporal evolution of the sources. Given the precise knowledge of the source distributions, the fields are now uniquely determined by solving Maxwell's macroscopic equations [2,4], namely,

$$\nabla \cdot D(r,t) = \rho_{\text{free}}(r,t), \qquad (3)$$

$$\nabla \times H(r,t) = J_{\text{free}}(r,t) + \partial D(r,t)/\partial t, \qquad (4)$$

$$\nabla \times E(r,t) = -\partial B(r,t)/\partial t, \qquad (5)$$

$$\nabla \cdot B(r,t) = 0. \qquad (6)$$

Here $\rho_{\text{free}}$ and $J_{\text{free}}$ are explicitly labeled as the densities of free charge and free current, $E$ is the electric field, $H$ is the magnetic field, $D$ is the displacement, and $B$ is the magnetic induction. By definition, $D(r,t) = \varepsilon_o E + P$ and $B(r,t) = \mu_o H + M$. The speed of light in vacuum is related to the EM properties of free space via $c = 1/\sqrt{\mu_o \varepsilon_o}$. Once again, $E$ and $B$ form a 2$^{nd}$ rank tensor that can be Lorentz transformed between inertial frames. A second such tensor is formed by $P$ and $M$, and a third one by $D$ and $H$. If we now define bound-charge and bound-current densities $\rho_{\text{bound}}(r,t) = -\nabla \cdot P$ and $J_{\text{bound}}(r,t) = (\partial P/\partial t) + \mu_o^{-1} \nabla \times M$, we may use them to eliminate $D$ and $H$ from Eqs.(3-6). Maxwell's macroscopic equations then reduce to his microscopic equations, where $E$ and $B$ are directly related to the total charge- and current-densities $\rho_{\text{total}} = \rho_{\text{free}} + \rho_{\text{bound}}$ and $J_{\text{total}} = J_{\text{free}} + J_{\text{bound}}$. Once again, $(J_{\text{total}}, c\rho_{\text{total}})$ acts as a 4-vector, and the continuity equation, $\nabla \cdot J_{\text{total}} + \partial \rho_{\text{total}}/\partial t = 0$, is automatically satisfied. It is thus clear that, as far as Maxwell's equations are concerned, $P(r,t)$ and $M(r,t)$ can be treated as distributions of charge and current, and that the knowledge of $\rho_{\text{total}}(\boldsymbol{r},t)$ and $\boldsymbol{J}_{\text{total}}(\boldsymbol{r},t)$ is all that is needed to determine the corresponding EM fields throughout the entire space-time.

The presence of magnetization in Maxwell's equations, however, creates difficulties with regard to energy and momentum conservation [5-9], and also causes trouble with the Lorentz force and torque equations that leads to apparent violations of special relativity under certain circumstances [10]. One way to avoid these difficulties is to resort to the four-vector formulation of the Lorentz force, or to assume the existence of a *hidden* energy flux $\mu_0^{-1}\boldsymbol{M}(\boldsymbol{r},t) \times \boldsymbol{E}(\boldsymbol{r},t)$ and also a *hidden* momentum-density $\varepsilon_0 \boldsymbol{M}(\boldsymbol{r},t) \times \boldsymbol{E}(\boldsymbol{r},t)$ wherever and whenever the $\boldsymbol{E}$ field is found to interact with magnetic dipoles [5-9,11-17]. An alternative way is to automatically remove the hidden entities from the relevant equations and to proceed by using these reduced equations without ever invoking the hidden entities [13,18,19]. Removing the hidden energy flux from the aforementioned Poynting vector yields the reduced Poynting vector $\boldsymbol{S}_2(\boldsymbol{r},t) = \boldsymbol{E} \times \boldsymbol{H}$ which is recognized as the standard form used in many textbooks [2-4]. The energy density of the EM field then becomes $\mathcal{E}_2(\boldsymbol{r},t) = \tfrac{1}{2}\varepsilon_0 \boldsymbol{E} \cdot \boldsymbol{E} + \tfrac{1}{2}\mu_0 \boldsymbol{H} \cdot \boldsymbol{H}$, and the energy exchange rate with the material subsystem becomes $\boldsymbol{E} \cdot \boldsymbol{J}_{\text{free}} + \boldsymbol{E} \cdot \partial \boldsymbol{P}/\partial t + \boldsymbol{H} \cdot \partial \boldsymbol{M}/\partial t$. Also, eliminating the hidden contribution to EM momentum yields the reduced momentum density $\boldsymbol{p}_{EM\_2}(\boldsymbol{r},t) = \boldsymbol{E} \times \boldsymbol{H}/c^2$, which is readily recognized as the Abraham momentum-density of the EM field [20-28]. Finally, removing from the Lorentz force expression the contribution of hidden momentum yields the following formula for the EM force-density under all circumstances:

$$\boldsymbol{F}_2(\boldsymbol{r},t) = (\rho_{\text{free}} - \nabla \cdot \boldsymbol{P})\boldsymbol{E} + [\boldsymbol{J}_{\text{free}} + (\partial \boldsymbol{P}/\partial t) + \mu_0^{-1}\nabla \times \boldsymbol{M}] \times \boldsymbol{B} - \partial(\varepsilon_0 \boldsymbol{M} \times \boldsymbol{E})/\partial t. \qquad (7)$$

Equation (7) can be shown [29] to be very similar (but not identical) to another equation first proposed by Einstein and Laub in 1908 [30,31]. The Einstein-Laub force-density equation is

$$\boldsymbol{F}_3(\boldsymbol{r},t) = \rho_{\text{free}}\boldsymbol{E} + \boldsymbol{J}_{\text{free}} \times \mu_0 \boldsymbol{H} + (\boldsymbol{P} \cdot \nabla)\boldsymbol{E} + (\partial \boldsymbol{P}/\partial t) \times \mu_0 \boldsymbol{H} + (\boldsymbol{M} \cdot \nabla)\boldsymbol{H} - (\partial \boldsymbol{M}/\partial t) \times \varepsilon_0 \boldsymbol{E}. \qquad (8)$$

When Eqs. (7) and (8) are integrated over the volume of an isolated object (i.e., an object surrounded by free space), the total force experienced by the object in the two formulations turns out to be the same [29]. In other words, Eqs. (7) and (8) may predict different force *distributions* within a given object, but the *total* forces predicted by them are identical. As for the torque density, in the case of Eq. (7), one will have $\boldsymbol{T}_2(\boldsymbol{r},t) = \boldsymbol{r} \times \boldsymbol{F}_2(\boldsymbol{r},t)$, whereas with Eq. (8) it becomes necessary to use the alternative expression

$$\boldsymbol{T}_3(\boldsymbol{r},t) = \boldsymbol{r} \times \boldsymbol{F}_3(\boldsymbol{r},t) + \boldsymbol{P} \times \boldsymbol{E} + \boldsymbol{M} \times \boldsymbol{H}. \qquad (9)$$

Once again, the two formulations yield identical *total* torques on any isolated object, but torque-density *distributions* are not necessarily the same [29]. The validity of neither formulation is ascertainable on theoretical grounds alone; rather, experimental tests are needed to determine which formulation leads to the observed force and torque *distributions* inside material media. It must be emphasized that field momenta and energy-momentum tensors *cannot* be measured directly. The *only* measurable entities are force and torque (and, with some effort in the case of deformable media, the local *densities* of force and torque) exerted by EM fields on material media. The various formulations of classical electrodynamics differ in their postulated forms of the energy-momentum tensor and/or the EM momentum density, but they predict the same *total* force and *total* torque on various objects. Whenever *total* force and *total* torque have been measured in the past, the results have been found to agree with apparently different theories when all the various ingredients of each theory are properly taken into consideration [32]. The differences between these theories, however, appear in their differing predictions of force and/or torque *distributions*, a subject that has been addressed in an elegant recent paper [33]. A main objective of the present paper is to show the differences in force- and torque-density *distributions* between the Einstein-Laub formulation and the Lorentz formulation–with the contribution of hidden momentum removed from the latter.

Einstein and Laub did not provide a firm basis for the force-density expression in Eq. (8). The closest that Einstein came to explaining his treatment of magnetization in the context of Maxwell's equations was the following statement [34]: "Now we have to fit [Maxwell's] equations … to the case where magnetically polarizable bodies are present. H. A. Lorentz does this by conceiving of certain electricities as being endowed with cyclical motions; from the standpoint of the pure electron theory this is also the only way that is justified. But for the sake of simplicity, we will base ourselves here on the knowledge that, as regards spatio-temporal interrelations, the magnetic polarization is a state wholly analogous to the polarization of dielectrics. Thus, we permit ourselves to conceive of magnetically polarizable bodies as being endowed with bound magnetic volume densities." In later years, Einstein came to believe that his conception had been incorrect, and regarded Minkowski's "strain tensor" as superior to his own [35]. It would take more than a decade after Einstein had passed away for Shockley [5-7] to discover the existence of "hidden momentum" in EM systems, thus revealing a major discrepancy between the Lorentz force and momentum conservation in situations involving an electric field acting on a magnetic material. As discussed in the preceding paragraphs, when the contribution of hidden momentum is eliminated from the Lorentz force, the resulting force-density expression comes tantalizingly close to that of Einstein and Laub. Similarly, the torque-density associated with the Einstein-Laub formulation comes close to that associated with the Lorentz law minus the hidden momentum contribution. It is, in fact, quite likely that the Einstein-Laub force and torque equations are the correct laws of classical electrodynamics that, in conjunction with Maxwell's macroscopic equations, should replace the Lorentz law. The present paper explores the similarities and differences of Eqs. (7) and (8) in the context of a few exemplary situations.

The following sections present examples of force and torque calculations using the Lorentz law (with and without the hidden momentum contribution), and also using the Einstein-Laub force and torque equations. These comparative studies also highlight areas where experiments can contribute to a better understanding of the two formulations. In the case of the Aharonov-Casher effect discussed in Sec. 2, it will be seen that the Lorentz formula predicts the existence of a force and torque on the magnetic dipole, whereas removing the contribution of hidden momentum brings the results into agreement with the predictions of the Einstein-Laub formula. Similarly, in Sec. 3, a magnetic dipole in the EM field of a moving charged-particle is seen to behave differently according to the Lorentz and Einstein-Laub formulations, but agreement is attained when the effect of hidden momentum is removed from the Lorentz force. In subsequent sections, several examples reveal the differences between the Lorentz law and the Einstein-Laub formulation in situations where the force-density *distribution* is of primary concern.

**2. The Aharonov-Casher effect**. A charge-neutral magnetic dipole moving in a static electric field does not experience any force or torque, yet its Schrödinger wave-function acquires a measurable phase, as was argued originally by Aharonov and Casher in a 1984 paper [36]. Boyer [37] objected to the claim of non-vanishing force on a moving magnetic dipole in a static *E*-field, but Aharonov *et al* [38] showed that the inclusion of hidden momentum in the analysis would confirm the original claim. In this section, we show that the predictions of the Einstein-Laub equations are in complete accord with the arguments of Aharonov and co-workers.

With reference to Fig. 1, consider a magnetic point-dipole $m_o\hat{x}$ located at $(x, y, z) = (0, 0, d)$ at $t = 0$, moving at a constant velocity $\boldsymbol{V} = V\hat{\boldsymbol{y}}$ in the $yz$-plane of a Cartesian coordinate system. Relativistic transformation from the rest frame of the dipole to the $xyz$ frame reveals the magnetization and polarization of the point-particle in the $xyz$ frame to be

$$\boldsymbol{M}(\boldsymbol{r},t) = \gamma m_o \hat{\boldsymbol{x}} \delta(x)\delta[\gamma(y-Vt)]\delta(z-d), \tag{10}$$

$$\boldsymbol{P}(\boldsymbol{r},t) = -\gamma V \varepsilon_o m_o \hat{\boldsymbol{z}} \delta(x)\delta[\gamma(y-Vt)]\delta(z-d). \tag{11}$$

In the above equations $\gamma = 1/\sqrt{1-(V/c)^2}$. We may proceed to drop $\gamma$ from both equations because $\gamma\delta[\gamma(y-Vt)] = \delta(y-Vt)$. It must also be pointed out that, in our notation, the magnetic moment of a loop of current $I_o$ having area $A$ is taken to be $m_o = \mu_o I_o A$ (as opposed to the conventional $m_o = I_o A$), the reason being that we have defined $\boldsymbol{M}$ such that $\boldsymbol{B} = \mu_o \boldsymbol{H} + \boldsymbol{M}$.

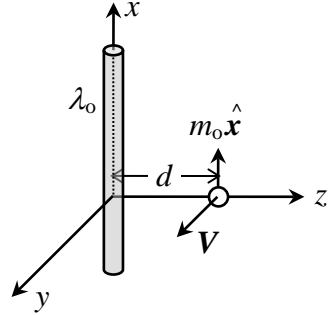

**Fig. 1**. Magnetic point-dipole $m_o\hat{x}$ moving at a constant velocity $V$ along the $y$-axis. At $t = 0$, the dipole is located at $(x, y, z) = (0, 0, d)$. Also present is a static electric field $\boldsymbol{E}(\boldsymbol{r}) = E_y(y,z)\hat{\boldsymbol{y}} + E_z(y,z)\hat{\boldsymbol{z}}$. This $E$-field could be produced, for example, by a long, straight, uniformly-charged wire along the $x$-axis.

The magnetic dipole is thus seen to be accompanied by a co-moving electric point-dipole $p_o\hat{\boldsymbol{z}}$, where $p_o = -\varepsilon_o m_o V$. Let us assume that a static electric field $\boldsymbol{E}(\boldsymbol{r}) = E_y(y,z)\hat{\boldsymbol{y}} + E_z(y,z)\hat{\boldsymbol{z}}$ is present in the $xyz$ frame. This field, which has no $x$-component, is also uniform along the $x$-axis. The conventional Lorentz force exerted on the moving dipole pair $(m_o\hat{\boldsymbol{x}}, p_o\hat{\boldsymbol{z}})$ is due to its bound electric charge and current densities, namely,

$$\rho_{\text{bound}}(\boldsymbol{r},t) = -\nabla\cdot\boldsymbol{P} = \varepsilon_o m_o V \delta(x)\delta(y-Vt)\delta'(z-d), \tag{12}$$

$$\boldsymbol{J}_{\text{bound}}(\boldsymbol{r},t) = \partial\boldsymbol{P}/\partial t + \mu_o^{-1}\nabla\times\boldsymbol{M}$$
$$= \mu_o^{-1} m_o \delta(x)\{\delta(y-Vt)\delta'(z-d)\hat{\boldsymbol{y}} - (1-V^2/c^2)\delta'(y-Vt)\delta(z-d)\hat{\boldsymbol{z}}\}. \tag{13}$$

In the absence of an external magnetic field, the Lorentz force-density $\boldsymbol{F}_1(\boldsymbol{r},t) = \rho\boldsymbol{E} + \boldsymbol{J}\times\boldsymbol{B}$ reduces to its first term. Integration over the relevant volume of space then yields the total force experienced by the moving dipole, as follows:

$$\boldsymbol{F}_{\text{total}}(t) = \int_{-\infty}^{\infty}\rho_{\text{bound}}(\boldsymbol{r},t)\boldsymbol{E}(\boldsymbol{r})\mathrm{d}x\mathrm{d}y\mathrm{d}z = -\varepsilon_o m_o V \frac{\partial\boldsymbol{E}(\boldsymbol{r})}{\partial z}\bigg|_{\boldsymbol{r}=(0,Vt,d)}. \tag{14}$$

This force, however, does not affect the motion of the dipole, as it is entirely used up by the time-rate-of-change of the hidden momentum $\boldsymbol{p}_{\text{hidden}}$, whose density is $\varepsilon_o \boldsymbol{M}(\boldsymbol{r},t) \times \boldsymbol{E}(\boldsymbol{r})$. To see this, observe that

$$\frac{\mathrm{d}\boldsymbol{p}_{\text{hidden}}}{\mathrm{d}t} = \frac{\mathrm{d}}{\mathrm{d}t}\int_{-\infty}^{\infty}\varepsilon_o \boldsymbol{M}(\boldsymbol{r},t)\times \boldsymbol{E}(\boldsymbol{r})\mathrm{d}x\mathrm{d}y\mathrm{d}z = \varepsilon_o m_o V\left(\frac{\partial E_y}{\partial y}\hat{\boldsymbol{z}} - \frac{\partial E_z}{\partial y}\hat{\boldsymbol{y}}\right)_{\boldsymbol{r}=(0,Vt,d)}. \tag{15}$$

The $E$-field being static, its divergence and curl must vanish, that is, $\partial E_y/\partial y + \partial E_z/\partial z = 0$ and $\partial E_z/\partial y = \partial E_y/\partial z$. Consequently, Eq. (15) may be written

$$\frac{\mathrm{d}\boldsymbol{p}_{\text{hidden}}}{\mathrm{d}t} = -\varepsilon_o m_o V \frac{\partial \boldsymbol{E}(0,Vt,d)}{\partial z}. \tag{16}$$

Comparison with Eq. (14) reveals that the force exerted by the $E$-field on the moving dipole is needed in its entirety to account for the time-rate-of-change of the hidden momentum. Similarly, one may compute the torque exerted on the moving dipole in accordance with the Lorentz law, as follows:

$$\boldsymbol{T}_{\text{total}}(t) = \int_{\text{all space}} \boldsymbol{r}\times \boldsymbol{F}_1(\boldsymbol{r},t)\mathrm{d}x\mathrm{d}y\mathrm{d}z = \int_{-\infty}^{\infty}\rho_{\text{bound}}(\boldsymbol{r},t)(yE_z - zE_y)\hat{\boldsymbol{x}}\,\mathrm{d}x\mathrm{d}y\mathrm{d}z$$

$$= V\varepsilon_o m_o \hat{\boldsymbol{x}}\left(-Vt\frac{\partial E_z}{\partial z} + d\frac{\partial E_y}{\partial z} + E_y\right)_{\boldsymbol{r}=(0,Vt,d)}$$

$$= (Vt\hat{\boldsymbol{y}} + d\hat{\boldsymbol{z}})\times \boldsymbol{F}_{\text{total}}(t) + \varepsilon_o m_o V\hat{\boldsymbol{x}}E_y(y=Vt,z=d). \tag{17}$$

The first term on the right-hand-side of Eq. (17) is the expected $\boldsymbol{r}\times \boldsymbol{F}$ contribution, which is related to the force exerted on the point-particle at $\boldsymbol{r} = Vt\hat{\boldsymbol{y}} + d\hat{\boldsymbol{z}}$. The second term arises from the action of the local $E$-field on the induced electric dipole, namely, $p_o\hat{\boldsymbol{z}}\times \boldsymbol{E}(\boldsymbol{r})$. As was the case with the Lorentz force in the preceding discussion, the torque in Eq. (17) is not real, in the sense that it does not cause a rotation of the point-particle around the $x$-axis; rather, it simply accounts for the time-rate-of-change of the hidden angular momentum $\boldsymbol{\mathcal{L}}_{\text{hidden}}$.

$$\frac{\mathrm{d}\boldsymbol{\mathcal{L}}_{\text{hidden}}}{\mathrm{d}t} = \frac{\mathrm{d}}{\mathrm{d}t}\int_{-\infty}^{\infty}\boldsymbol{r}\times[\varepsilon_o \boldsymbol{M}(\boldsymbol{r},t)\times \boldsymbol{E}(\boldsymbol{r})]\mathrm{d}x\mathrm{d}y\mathrm{d}z$$

$$= \varepsilon_o m_o V\hat{\boldsymbol{x}}\left(E_y + Vt\frac{\partial E_y}{\partial y} + d\frac{\partial E_z}{\partial y}\right) = \varepsilon_o m_o V\hat{\boldsymbol{x}}\left(E_y - Vt\frac{\partial E_z}{\partial z} + d\frac{\partial E_y}{\partial z}\right)_{\boldsymbol{r}=(0,Vt,d)}$$

$$= (Vt\hat{\boldsymbol{y}} + d\hat{\boldsymbol{z}})\times \boldsymbol{F}_{\text{total}}(t) + \varepsilon_o m_o VE_y(y=Vt,z=d)\hat{\boldsymbol{x}}. \tag{18}$$

Next, we use the Einstein-Laub equations to derive the force and torque exerted by the $E$-field on the moving point-particle. For the force-density we find

$$\boldsymbol{F}_3(\boldsymbol{r},t) = (\boldsymbol{P}\cdot\nabla)\boldsymbol{E} - (\partial \boldsymbol{M}/\partial t)\times \varepsilon_o \boldsymbol{E}$$

$$= -\varepsilon_o m_o V\delta(x)\delta(y-Vt)\delta(z-d)(\partial \boldsymbol{E}/\partial z) + \varepsilon_o m_o V\delta(x)\delta'(y-Vt)\delta(z-d)(E_y\hat{\boldsymbol{z}} - E_z\hat{\boldsymbol{y}}). \tag{19}$$

Integrating the above force-density over the relevant volume of space yields

$$\boldsymbol{F}_{\text{total}}(t) = \int_{-\infty}^{\infty} \boldsymbol{F}_3(\boldsymbol{r},t) \mathrm{d}x\mathrm{d}y\mathrm{d}z = \varepsilon_\mathrm{o} m_\mathrm{o} V \left[ -\frac{\partial \boldsymbol{E}}{\partial z} + \left( \frac{\partial E_z}{\partial y} \hat{\boldsymbol{y}} - \frac{\partial E_y}{\partial y} \hat{\boldsymbol{z}} \right) \right]_{\boldsymbol{r}=(0,Vt,d)}$$

$$= \varepsilon_\mathrm{o} m_\mathrm{o} V \left[ \left( \frac{\partial E_z}{\partial y} - \frac{\partial E_y}{\partial z} \right) \hat{\boldsymbol{y}} - \left( \frac{\partial E_y}{\partial y} + \frac{\partial E_z}{\partial z} \right) \hat{\boldsymbol{z}} \right]_{\boldsymbol{r}=(0,Vt,d)} = 0. \qquad (20)$$

In the above derivation, we have used, once again, the fact that, for a static electric field, $\nabla \cdot \boldsymbol{E}(\boldsymbol{r}) = 0$ and $\nabla \times \boldsymbol{E}(\boldsymbol{r}) = 0$. The Einstein-Laub force on the moving particle thus vanishes *without* the need to invoke hidden entities. The total torque on the moving particle in accordance with the Einstein-Laub formalism is similarly found to be zero, that is,

$$\boldsymbol{T}_{\text{total}}(t) = \int_{\text{all space}} [\boldsymbol{r} \times \boldsymbol{F}_3(\boldsymbol{r},t) + \boldsymbol{P}(\boldsymbol{r},t) \times \boldsymbol{E}(\boldsymbol{r},t) + \boldsymbol{M}(\boldsymbol{r},t) \times \boldsymbol{H}(\boldsymbol{r},t)] \mathrm{d}x\mathrm{d}y\mathrm{d}z$$

$$= -\varepsilon_\mathrm{o} m_\mathrm{o} V \hat{\boldsymbol{x}} \left[ \left( Vt \frac{\partial E_z}{\partial z} - d \frac{\partial E_y}{\partial z} \right) + \left( E_y + Vt \frac{\partial E_y}{\partial y} + d \frac{\partial E_z}{\partial y} \right) - E_y \right]_{\boldsymbol{r}=(0,Vt,d)} = 0. \qquad (21)$$

Clearly, the Einstein-Laub equations yield the correct force and torque on the moving dipole without assistance from hidden entities.

We close this section by recalling the essential idea behind the Aharonov-Casher effect. A charge-neutral magnetic dipole $\boldsymbol{m}_\mathrm{o}$ moving with velocity $\boldsymbol{V}$ produces a co-moving electric dipole $\boldsymbol{p}_\mathrm{o} = \varepsilon_\mathrm{o} \boldsymbol{V} \times \boldsymbol{m}_\mathrm{o}$. Denoting the mass of the particle by $\underset{\sim}{m}$, the (non-relativistic) Lagrangian of the dipole in the static electric field $\boldsymbol{E}(\boldsymbol{r})$ is given by

$$L(\boldsymbol{r},\boldsymbol{V},t) = \tfrac{1}{2}\underset{\sim}{m} V^2 + \boldsymbol{p}_\mathrm{o} \cdot \boldsymbol{E}(\boldsymbol{r}) = \tfrac{1}{2}\underset{\sim}{m} V^2 + \varepsilon_\mathrm{o}(\boldsymbol{V} \times \boldsymbol{m}_\mathrm{o}) \cdot \boldsymbol{E}(\boldsymbol{r}) = \tfrac{1}{2}\underset{\sim}{m} V^2 - \varepsilon_\mathrm{o} \boldsymbol{V} \cdot (\boldsymbol{E} \times \boldsymbol{m}_\mathrm{o}). \qquad (22)$$

With reference to Fig. 1, let the *E*-field be produced by a long, straight, uniformly-charged wire along the *x*-axis. Denoting the linear charge-density of the wire by $\lambda_\mathrm{o}$, its *E*-field will be $\boldsymbol{E}(\rho, \phi, z) = \lambda_\mathrm{o} \hat{\boldsymbol{\rho}}/(2\pi\varepsilon_\mathrm{o}\rho)$. While traversing a closed path in the *yz*-plane that encloses the *x*-axis, the dipole's quantum mechanical wave-function accumulates a phase $\Delta\phi$, despite the fact that no forces act on the particle in its entire path. The round-trip phase is given by

$$\Delta\phi = \frac{1}{\hbar} \int_{t_1}^{t_2} \varepsilon_\mathrm{o} (\boldsymbol{V} \times \boldsymbol{m}_\mathrm{o}) \cdot \boldsymbol{E}(\boldsymbol{r}) \mathrm{d}t = \frac{\varepsilon_\mathrm{o}}{\hbar} \oint_{\substack{\text{closed path in}\\ \text{the }yz\text{-plane}}} (\mathrm{d}\boldsymbol{r} \times \boldsymbol{m}_\mathrm{o}) \cdot \boldsymbol{E}(\boldsymbol{r}) = \frac{\varepsilon_\mathrm{o} \boldsymbol{m}_\mathrm{o}}{\hbar} \cdot \oint \boldsymbol{E}(\boldsymbol{r}) \times \mathrm{d}\boldsymbol{r}$$

$$= \frac{\varepsilon_\mathrm{o} m_\mathrm{o} \hat{\boldsymbol{z}}}{\hbar} \cdot \oint E_\rho(\rho) \hat{\boldsymbol{\rho}} \times (\hat{\boldsymbol{\rho}} \mathrm{d}\rho + \hat{\boldsymbol{\phi}} \rho \mathrm{d}\phi) = \frac{\varepsilon_\mathrm{o} m_\mathrm{o} \hat{\boldsymbol{z}}}{\hbar} \cdot \int_0^{2\pi} \hat{\boldsymbol{z}} E_\rho(\rho) \rho \, \mathrm{d}\phi = \frac{m_\mathrm{o} \lambda_\mathrm{o}}{\hbar}. \qquad (23)$$

This phenomenon is analogous to the Aharonov-Bohm effect [39], where a charged particle acquires a phase in traveling around a uniformly-magnetized wire (or solenoid), despite the fact that no forces are exerted on the particle by the vector potential field $\boldsymbol{A}(\boldsymbol{r})$ surrounding the wire. In the case of the Aharonov-Bohm effect, of course, there is no need to invoke the Einstein-Laub formulation, as $\boldsymbol{E}$, $\boldsymbol{B}$, and $\boldsymbol{H}$ fields are all zero in the region where the particle resides – only the vector potential $\boldsymbol{A}$ is nonzero in the space surrounding an infinitely-long solenoid. The preceding statement is true even when the point-charge is accompanied by an intrinsic magnetic moment (e.g., an electron, whose spin angular momentum produces its magnetic dipole moment).

**3. Magnetic dipole in a time-varying electric field.** Figure 2 shows a stationary magnetic point-dipole $m_o\hat{x}$ located at $(x, y, z) = (0, 0, d)$, whose magnetization may be written as follows:

$$M(r,t) = m_o\hat{x}\,\delta(x)\delta(y)\delta(z-d). \tag{24}$$

A point-charge $q$, moving along the $z$-axis and toward the point-dipole at a constant velocity $V = V\hat{z}$, passes through the origin of the coordinate system at $t = 0$. The electromagnetic field of the point-charge is readily obtained by a Lorentz transformation from the rest frame of the charge to the $xyz$ frame. We find

$$E(r,t) = \frac{\gamma q[x\hat{x} + y\hat{y} + (z-Vt)\hat{z}]}{4\pi\varepsilon_o[x^2 + y^2 + \gamma^2(z-Vt)^2]^{3/2}}, \tag{25a}$$

$$H(r,t) = \frac{\gamma Vq(x\hat{y} - y\hat{x})}{4\pi[x^2 + y^2 + \gamma^2(z-Vt)^2]^{3/2}}. \tag{25b}$$

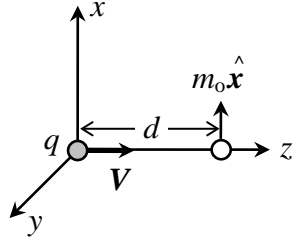

**Fig. 2**. A stationary magnetic dipole $m_o\hat{x}$ located at $r = (0, 0, d)$ experiences both the electric and the magnetic field of a point-charge $q$ moving along the $z$-axis at a constant velocity $V$.

The Lorentz force exerted by the moving point-charge on the stationary point-dipole is thus given by

$$F_1(t) = \int_{-\infty}^{\infty}[\mu_o^{-1}\nabla\times M(r,t)]\times \mu_o H(r,t)\,\mathrm{d}x\mathrm{d}y\mathrm{d}z$$

$$= \frac{m_o\gamma Vq}{4\pi}\int_{-\infty}^{\infty}\frac{\delta(x)[\delta(y)\delta'(z-d)\hat{y} - \delta'(y)\delta(z-d)\hat{z}]\times(x\hat{y} - y\hat{x})}{[x^2 + y^2 + \gamma^2(z-Vt)^2]^{3/2}}\,\mathrm{d}x\mathrm{d}y\mathrm{d}z$$

$$= \frac{m_o\gamma Vq\,\hat{y}}{4\pi}\int_{-\infty}^{\infty}\frac{y\delta'(y)}{[y^2 + \gamma^2(d-Vt)^2]^{3/2}}\,\mathrm{d}y = -\frac{m_o Vq\,\hat{y}}{4\pi\gamma^2|d-Vt|^3}. \tag{26}$$

In contrast, the Einstein-Laub formula predicts the following force:

$$F_3(t) = \int_{-\infty}^{\infty}[(M\cdot\nabla)H - (\partial M/\partial t)\times\varepsilon_o E]\,\mathrm{d}x\mathrm{d}y\mathrm{d}z$$

$$= \int_{-\infty}^{\infty}m_o\delta(x)\delta(y)\delta(z-d)\frac{\partial}{\partial x}\left\{\frac{\gamma Vq(x\hat{y} - y\hat{x})}{4\pi[x^2 + y^2 + \gamma^2(z-Vt)^2]^{3/2}}\right\}\mathrm{d}x\mathrm{d}y\mathrm{d}z$$

$$= \frac{m_o Vq\,\hat{y}}{4\pi\gamma^2|d-Vt|^3}. \tag{27}$$

This force is equal in magnitude but opposite in direction to the Lorentz force of Eq. (26). The discrepancy is resolved when the hidden momentum contribution is taken into account. We will have

$$\frac{\mathrm{d}\boldsymbol{p}_{\text{hidden}}}{\mathrm{d}t} = \frac{\mathrm{d}}{\mathrm{d}t}\int_{-\infty}^{\infty}\varepsilon_{\text{o}}\boldsymbol{M}(\boldsymbol{r},t)\times\boldsymbol{E}(\boldsymbol{r},t)\mathrm{d}x\mathrm{d}y\mathrm{d}z$$

$$= \frac{\mathrm{d}}{\mathrm{d}t}\int_{-\infty}^{\infty}\varepsilon_{\text{o}}m_{\text{o}}\delta(x)\delta(y)\delta(z-d)\hat{\boldsymbol{x}}\times\frac{\gamma q[x\hat{\boldsymbol{x}}+y\hat{\boldsymbol{y}}+(z-Vt)\hat{\boldsymbol{z}}]}{4\pi\varepsilon_{\text{o}}[x^2+y^2+\gamma^2(z-Vt)^2]^{3/2}}\mathrm{d}x\mathrm{d}y\mathrm{d}z$$

$$= -\frac{m_{\text{o}}q\hat{\boldsymbol{y}}}{4\pi\gamma^2}\mathrm{sign}(d-Vt)\frac{\mathrm{d}}{\mathrm{d}t}\left\{\frac{1}{(d-Vt)^2}\right\} = -\frac{2m_{\text{o}}Vq\hat{\boldsymbol{y}}}{4\pi\gamma^2|d-Vt|^3}. \tag{28}$$

Here we have an interesting situation, where the Lorentz law predicts a force on the dipole along the negative $y$-axis, but the contribution of the hidden momentum compels the dipole to move in the opposite direction. The Einstein-Laub formula, however, predicts the correct magnitude and direction of force without invoking hidden entities.

The Lorentz formula also predicts the following torque exerted by the moving point-charge on the point-dipole:

$$\boldsymbol{T}_1(t) = \int_{-\infty}^{\infty}\boldsymbol{r}\times\left\{[\mu_{\text{o}}^{-1}\nabla\times\boldsymbol{M}(\boldsymbol{r},t)]\times\mu_{\text{o}}\boldsymbol{H}(\boldsymbol{r},t)\right\}\mathrm{d}x\mathrm{d}y\mathrm{d}z = \frac{m_{\text{o}}Vqd\hat{\boldsymbol{x}}}{4\pi\gamma^2|d-Vt|^3}. \tag{29}$$

This is simply the torque associated with the force in Eq.(26) acting at a distance $z=d$ from the origin. The corresponding Einstein-Laub torque is given by

$$\boldsymbol{T}_3(t) = \int_{-\infty}^{\infty}\left\{\boldsymbol{r}\times[(\boldsymbol{M}\cdot\nabla)\boldsymbol{H}-(\partial\boldsymbol{M}/\partial t)\times\varepsilon_{\text{o}}\boldsymbol{E}]+\boldsymbol{M}\times\boldsymbol{H}\right\}\mathrm{d}x\mathrm{d}y\mathrm{d}z = -\frac{m_{\text{o}}Vqd\hat{\boldsymbol{x}}}{4\pi\gamma^2|d-Vt|^3}. \tag{30}$$

Once again, the Einstein-Laub torque is seen to be due to the corresponding force, given by Eq.(27), acting at a distance $z=d$ from the origin. The Einstein-Laub and Lorentz torques are thus equal in magnitude but opposite in orientation. To explain this difference, we must look for the hidden angular momentum, namely,

$$\frac{\mathrm{d}\boldsymbol{\mathcal{L}}_{\text{hidden}}}{\mathrm{d}t} = \frac{\mathrm{d}}{\mathrm{d}t}\int_{-\infty}^{\infty}\boldsymbol{r}\times\left[\varepsilon_{\text{o}}\boldsymbol{M}(\boldsymbol{r},t)\times\boldsymbol{E}(\boldsymbol{r},t)\right]\mathrm{d}x\mathrm{d}y\mathrm{d}z = \frac{2m_{\text{o}}Vqd\hat{\boldsymbol{x}}}{4\pi\gamma^2|d-Vt|^3}. \tag{31}$$

The difference between the Lorentz torque and the Einstein-Laub torque is thus seen to be entirely due to the hidden angular momentum.

**4. Observable differences between the Lorentz force and the Einstein-Laub force**. With the hidden momentum contribution removed from the Lorentz force, one can show that the Lorentz and the Einstein-Laub formulas always yield the same *total* force (and also *total* torque) on any given object exposed to EM fields [29]. However, the force and torque *distributions* predicted by the two formulas are often quite different, and here lies an opportunity for experiments to test the veracity of either formulation. Consider, for instance, a thin, uniformly-polarized rod of length $2L$ parallel to the $x$-axis at a distance $d$ from a point charge $q$, as shown in Fig.3. The polarization of the rod may be written as

$$\boldsymbol{P}(\boldsymbol{r}) = P_{\text{o}}\mathrm{Rect}(x/2L)\delta(y)\delta(z-d)\hat{\boldsymbol{x}}. \tag{32}$$

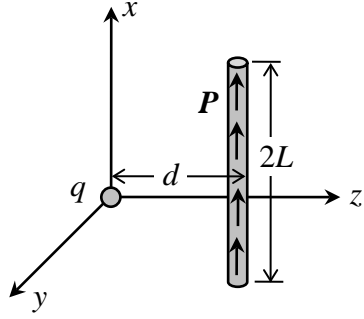

**Fig. 3**. A point charge $q$, sitting at the origin of the coordinate system, exerts a force and a torque on a thin, straight rod of a uniformly-polarized material. The center of the rod is at $\mathbf{r} = (0,0,d)$.

Here the rectangle function $\text{Rect}(x)$ is equal to 1.0 when $|x| \leq \frac{1}{2}$, and zero otherwise. The Lorentz force-density experienced by the rod in the presence of the point-charge $q$ located at the origin of the coordinate system is

$$\mathbf{F}_1(\mathbf{r}) = -[\nabla \cdot \mathbf{P}(\mathbf{r})]\mathbf{E}(\mathbf{r}) = \frac{qP_o[(L\hat{\mathbf{x}}+d\hat{\mathbf{z}})\delta(x-L)+(L\hat{\mathbf{x}}-d\hat{\mathbf{z}})\delta(x+L)]\delta(y)\delta(z-d)}{4\pi\varepsilon_o(L^2+d^2)^{3/2}}. \quad (33)$$

The Lorentz force is thus seen to be concentrated at the top and bottom of the rod, with no effect whatsoever along its length. In contrast, the Einstein-Laub force-density is given by

$$\mathbf{F}_3(\mathbf{r}) = [\mathbf{P}(\mathbf{r})\cdot\nabla]\mathbf{E}(\mathbf{r}) = \frac{qP_o[(d^2-2x^2)\hat{\mathbf{x}}-3dx\hat{\mathbf{z}}]\text{Rect}(x/2L)\delta(y)\delta(z-d)}{4\pi\varepsilon_o(x^2+d^2)^{5/2}}. \quad (34)$$

This force distribution, which varies continuously with position along $x$, must stress the rod differently than the Lorentz force-density of Eq. (33) and should, therefore, be measurable. The component of total force along the $z$-axis is readily seen to vanish because $F_z$ is an odd function of $x$. The total force, which is aligned with the $x$-axis, may be found by direct integration of Eq. (34) as follows:

$$\int_{-\infty}^{\infty} F_x(\mathbf{r})\,dxdydz = \frac{qP_o}{4\pi\varepsilon_o}\int_{-L}^{+L}\frac{d^2-2x^2}{(x^2+d^2)^{5/2}}dx = \left.\frac{qP_ox}{4\pi\varepsilon_o(x^2+d^2)^{3/2}}\right|_{-L}^{+L} = \frac{qP_oL}{2\pi\varepsilon_o(L^2+d^2)^{3/2}}. \quad (35)$$

This is identical with the total Lorentz force on the rod, obtained from Eq. (33). In similar fashion, one can readily confirm the equality of total torques calculated in accordance with the two formulations.

**5. Magnetic rod in an external magnetic field**. Consider a thin, uniformly-magnetized rod of length $2L$ placed parallel to the $x$-axis at $\mathbf{r} = (0,0,d)$, as shown in Fig. 4. The magnetization of the rod may be described by the following function:

$$\mathbf{M}(\mathbf{r}) = M_o\,\text{Rect}(x/2L)\delta(y)\delta(z-d)\hat{\mathbf{x}}. \quad (36)$$

The external field acting on the above rod is produced by a constant current in a long, thin, straight wire placed at the origin and aligned with the $y$-axis. The $H$-field is thus given by

$$\mathbf{H}(\mathbf{r}) = \frac{I_o(z\hat{\mathbf{x}}-x\hat{\mathbf{z}})}{2\pi(x^2+z^2)}. \quad (37)$$

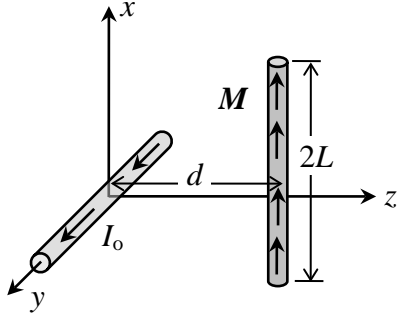

**Fig. 4**. A long, thin, straight wire carrying a constant current $I_o$ along the $y$-axis produces a magnetic field $\boldsymbol{H}(x,z)$ that exerts force and torque on a thin, straight, uniformly-magnetized rod. The rod, which is aligned with the $x$-axis, is centered at $\boldsymbol{r} = (0,0,d)$.

The Lorentz force-density acting on the magnetized rod is calculated as follows:

$$\boldsymbol{F}_1(x) = \int_{-\infty}^{\infty} [\mu_o^{-1}\nabla \times \boldsymbol{M}(\boldsymbol{r})] \times \mu_o \boldsymbol{H}(\boldsymbol{r}) \mathrm{d}y\mathrm{d}z = -\frac{I_o M_o [2dx\hat{\boldsymbol{x}} + (d^2 - x^2)\hat{\boldsymbol{z}}]\mathrm{Rect}(x/2L)}{2\pi(d^2 + x^2)^2}. \tag{38}$$

The total force is now obtained by integrating $\boldsymbol{F}_1(x)$ along the length of the rod, that is,

$$\int_{-\infty}^{\infty} \boldsymbol{F}_1(x)\mathrm{d}x = -\frac{I_o M_o L\hat{\boldsymbol{z}}}{\pi(L^2 + d^2)}. \tag{39}$$

As for the force-density according to the Einstein-Laub formula, we have

$$\boldsymbol{F}_3(x) = \int_{-\infty}^{\infty} [\boldsymbol{M}(\boldsymbol{r})\cdot\nabla]\boldsymbol{H}(\boldsymbol{r})\mathrm{d}y\mathrm{d}z = -\frac{I_o M_o [2dx\hat{\boldsymbol{x}} + (d^2 - x^2)\hat{\boldsymbol{z}}]\mathrm{Rect}(x/2L)}{2\pi(d^2 + x^2)^2}. \tag{40}$$

The two formulations thus yield identical results for the force distribution along the length of the magnetized rod. The torque may also be calculated using similar procedures, and the net torque with respect to the origin of coordinates is found to vanish in both formulations. Unlike the example of the preceding section, where a polarized rod in an external $E$-field exhibited different force-density distributions according to the Lorentz and Einstein-Laub schemes, a magnetized rod in an external $H$-field is seen to be an inadequate vehicle for comparing the two formulations.

**6. Magnetized cylinder in a uniform external magnetic field**. As a simple example involving a magnetic material that *can* be used to distinguish between the Lorentz and Einstein-Laub formulations, consider a solid cylinder of radius $R$ and length $2L$, uniformly magnetized along the $z$-axis, as shown in Fig. 5. The magnetization of the cylinder is thus written

$$\boldsymbol{M}(\boldsymbol{r}) = M_o\hat{\boldsymbol{z}}\,\mathrm{Circ}(\rho/R)\mathrm{Rect}(z/2L). \tag{41}$$

In the cylindrical coordinate system $(\rho, \phi, z)$, the function $\mathrm{Circ}(\rho)$ is defined as 1.0 when $\rho \le 1.0$, and 0 otherwise. Suppose the above cylinder is subjected to a constant, uniform magnetic field along the $z$-axis, that is, $\boldsymbol{H}(\boldsymbol{r}) = H_o\hat{\boldsymbol{z}}$. The Einstein-Laub formula yields zero force-density throughout the cylinder. [One may be inclined to argue that the cylinder is subject to a stretching force along the $z$-axis due to the action of the $H$-field on its constituent dipoles, but such stretching is not predicted by the Einstein-Laub equation $\boldsymbol{F}_3 = (\boldsymbol{M}\cdot\nabla)\boldsymbol{H}$.] In contrast, the Lorentz law predicts the existence of an expansive force with an areal density of $M_oH_o\hat{\boldsymbol{\rho}}$ acting on the cylindrical surface. The difference between these two situations is significant and should be amenable to experimental verification.

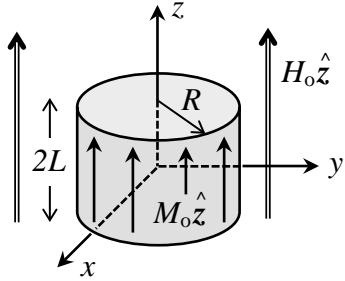

**Fig. 5**. A uniformly-magnetized solid cylinder of radius $R$, length $2L$, and magnetization $M_o\hat{\boldsymbol{z}}$, is subject to a constant, uniform magnetic field $\boldsymbol{H} = H_o\hat{\boldsymbol{z}}$.

**7. Magnetized ring in an external magnetic field**. With reference to Fig. 6, let an azimuthally-magnetized ring be subject to the magnetic field of a long, straight wire carrying a constant current $I_o$ along the ring's axis. We have

$$\boldsymbol{H}(\boldsymbol{r}) = \frac{I_o\hat{\boldsymbol{\phi}}}{2\pi\rho}. \tag{42}$$

$$\boldsymbol{M}(\boldsymbol{r}) = M_o\hat{\boldsymbol{\phi}}[\text{Circ}(\rho/R_2) - \text{Circ}(\rho/R_1)]\,\text{Rect}(z/2L). \tag{43}$$

The Lorentz force-density is given by

$$\boldsymbol{F}_1(\boldsymbol{r}) = [\mu_o^{-1}\boldsymbol{\nabla}\times\boldsymbol{M}(\boldsymbol{r})]\times\mu_o\boldsymbol{H}(\boldsymbol{r}) = \left[-\frac{\partial M_\phi}{\partial z}\hat{\boldsymbol{\rho}} + \frac{\partial(\rho M_\phi)}{\rho\partial\rho}\hat{\boldsymbol{z}}\right]\times\frac{I_o\hat{\boldsymbol{\phi}}}{2\pi\rho}$$

$$= \frac{I_oM_o}{2\pi\rho}\Big\{\{\delta(\rho-R_2)-\delta(\rho-R_1)-\rho^{-1}[\text{Circ}(\rho/R_2)-\text{Circ}(\rho/R_1)]\}\text{Rect}(z/2L)\hat{\boldsymbol{\rho}}$$

$$+ [\text{Circ}(\rho/R_2)-\text{Circ}(\rho/R_1)][\delta(z-L)-\delta(z+L)]\hat{\boldsymbol{z}}\Big\}. \tag{44}$$

In the above expression, the four delta-functions represent surface forces that tend to expand the ring by pulling on its four surfaces. In addition, there is the volumetric force density, $-(I_oM_o/2\pi\rho^2)\hat{\boldsymbol{\rho}}$, which tends to stretch the ring radially toward the $z$-axis. The net force and torque on the ring are seen to be zero, but the distribution of force within the ring and on its surfaces should be accessible to experiments.

Next, we calculate the Einstein-Laub force-density exerted by the external $H$-field on the magnetic material of the ring. We find

$$\boldsymbol{F}_3(\boldsymbol{r}) = (\boldsymbol{M}\cdot\boldsymbol{\nabla})\boldsymbol{H} = M_\rho\frac{\partial\boldsymbol{H}}{\partial\rho} + M_\phi\frac{\partial\boldsymbol{H}}{\rho\partial\phi} + M_z\frac{\partial\boldsymbol{H}}{\partial z}$$

$$= M_\rho\frac{\partial\boldsymbol{H}}{\partial\rho} + \frac{M_\phi}{\rho}\left[\frac{\partial H_\rho}{\partial\phi}\hat{\boldsymbol{\rho}} + \frac{\partial H_\phi}{\partial\phi}\hat{\boldsymbol{\phi}} + \frac{\partial H_z}{\partial\phi}\hat{\boldsymbol{z}} + H_\rho\hat{\boldsymbol{\phi}} - H_\phi\hat{\boldsymbol{\rho}}\right] + M_z\frac{\partial\boldsymbol{H}}{\partial z}$$

$$= -\frac{I_oM_o[\text{Circ}(\rho/R_2)-\text{Circ}(\rho/R_1)]\,\text{Rect}(z/2L)\hat{\boldsymbol{\rho}}}{2\pi\rho^2}. \tag{45}$$

This volumetric force-density is seen to be the same as that obtained with the Lorentz law in Eq.(44), but the forces acting on the ring's surfaces are now absent. An experiment designed to detect the presence of surface forces should thus be able to decide between the Einstein-Laub and Lorentz formulations.

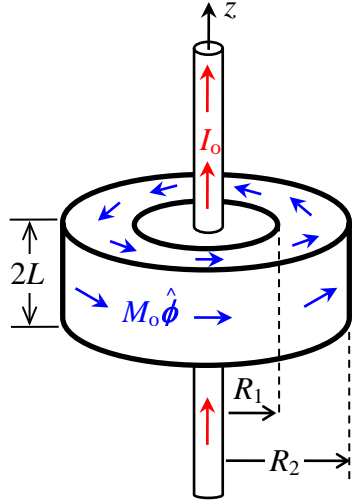

**Fig.6**. A permanently magnetized ring has inner radius $R_1$, outer radius $R_2$, length $2L$, and uniform magnetization $M_o\hat{\phi}$. The ring is subject to an external magnetic field produced by a long, straight wire that passes through the center of the ring and carries a constant, uniform current $I_o$ along the $z$-axis.

**8. Magnetized ring carrying a radially-directed electric current**. A disk of thickness $2L$, inner radius $R_1$, and outer radius $R_2$ has uniform magnetization $M_\phi(\rho)\hat{\phi}$. in the azimuthal direction. A constant radial current, having density $\boldsymbol{J}_{\text{free}}(\boldsymbol{r}) = J_o\hat{\boldsymbol{\rho}}/\rho$, flows from the inner to the outer cylindrical wall, as shown in Fig.7. In what follows, we shall determine the EM force-density experienced by the disk according to the Lorentz law as well as that due to the Einstein-Laub law. In both cases the total force on the disk will turn out to be zero. We mention in passing that Einstein [40] used this example to argue that the ponderomotive force-density on $\boldsymbol{J}_{\text{free}}$ should be given by $\boldsymbol{J}_{\text{free}} \times \mu_o\boldsymbol{H}$ as opposed to $\boldsymbol{J}_{\text{free}} \times \boldsymbol{B}$. He does not seem to have realized that, in the Lorentz formulation of this problem, the correct force-density is $(\boldsymbol{J}_{\text{free}} + \mu_o^{-1}\boldsymbol{\nabla} \times \boldsymbol{M}) \times \boldsymbol{B}$, which yields the same *total* force as the Einstein-Laub force-density of $\boldsymbol{J}_{\text{free}} \times \mu_o\boldsymbol{H} + (\boldsymbol{M} \cdot \boldsymbol{\nabla})\boldsymbol{H}$.

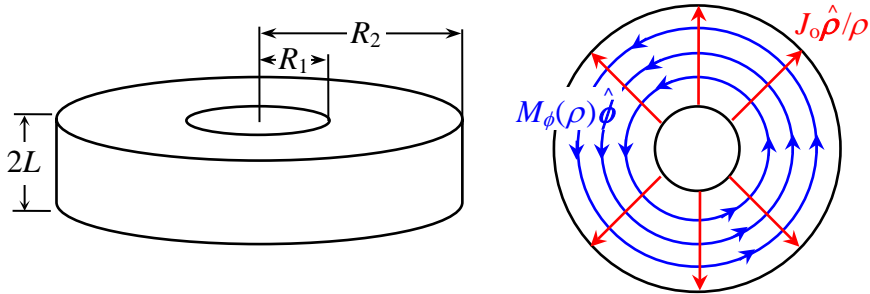

**Fig.7**. A permanently magnetized ring having inner radius $R_1$, outer radius $R_2$, length $2L$, and azimuthal magnetization $M_\phi(\rho)\hat{\phi}$ carries a constant radial current $J_o\hat{\boldsymbol{\rho}}/\rho$ between its inner and outer cylindrical walls.

The *H*-field produced by the magnetization of the disk is zero everywhere. This is because the only source of **H** in a magnetostatic system is the magnetic bound charge-density $\rho_{\text{bound}}^{(m)} = -\nabla \cdot \boldsymbol{M}(\boldsymbol{r})$, which is zero in the present problem. In the absence of an *H*-field, the only contribution to the disk's *B*-field comes from $\boldsymbol{M}(\boldsymbol{r})$, that is,

$$\boldsymbol{B}(\rho,\phi,z) = M_\phi(\rho)[\text{Circ}(\rho/R_2) - \text{Circ}(\rho/R_1)]\,\text{Rect}(z/2L)\hat{\boldsymbol{\phi}}. \tag{46}$$

The Lorentz force-density acting on the free current is thus given by

$$\boldsymbol{F}_1(\rho,z) = \boldsymbol{J}_{\text{free}} \times \boldsymbol{B} = \rho^{-1} J_o M_\phi(\rho)\,[\text{Circ}(\rho/R_2) - \text{Circ}(\rho/R_1)]\,\text{Rect}(z/2L)\hat{\boldsymbol{z}}. \tag{47}$$

Integrating the above force-density over the volume of the disk yields

$$\int_{\rho=R_1}^{R_2}\int_{z=-L}^{L} 2\pi\rho \boldsymbol{F}_1(\rho,z)\mathrm{d}\rho\mathrm{d}z = 4\pi L J_o \hat{\boldsymbol{z}} \int_{R_1}^{R_2} M_\phi(\rho)\,\mathrm{d}\rho. \tag{48}$$

The radial current produces its own magnetic field inside as well as outside the disk. Symmetry considerations indicate that the vector potential associated with $J_o\hat{\boldsymbol{\rho}}/\rho$ is radial and independent of $\phi$, that is, $\boldsymbol{A}(\boldsymbol{r}) = A(\rho,z)\hat{\boldsymbol{\rho}}$. Moreover, $A(\rho,z)$ must be an even function of *z*. Thus the magnetic field obtained using $\boldsymbol{B}(\boldsymbol{r}) = \nabla \times \boldsymbol{A}(\boldsymbol{r})$ is aligned with $\hat{\boldsymbol{\phi}}$ and is independent of $\phi$, that is, $\boldsymbol{H}(\boldsymbol{r}) = H(\rho,z)\hat{\boldsymbol{\phi}}$; it is also an odd function of *z*. Applying Ampere's law, $\nabla \times \boldsymbol{H} = \boldsymbol{J}_{\text{free}}$, to a small rectangular loop located inside the disk and oriented perpendicularly to $\hat{\boldsymbol{\rho}}$ yields

$$\boldsymbol{H}(\rho, z = \pm L) = \mp (J_o L/\rho)\,\hat{\boldsymbol{\phi}}; \qquad R_1 \leq \rho \leq R_2. \tag{49}$$

The odd symmetry of the above *H*-field with respect to *z* means that the force it exerts on $\boldsymbol{J}_{\text{free}}$, when integrated along *z*, vanishes. The other current in the system is the bound electric current of $\boldsymbol{M}(\boldsymbol{r})$, which has both radial and vertical components, as follows:

$$\boldsymbol{J}_{\text{bound}} = \mu_o^{-1}\nabla \times \boldsymbol{M}(\boldsymbol{r}) = \mu_o^{-1}\{M_\phi(\rho)[\delta(z-L) - \delta(z+L)]\hat{\boldsymbol{\rho}} + [\rho^{-1}M_\phi(\rho) + M'_\phi(\rho)$$
$$+ M_\phi(R_1)\delta(\rho - R_1) - M_\phi(R_2)\delta(\rho - R_2)]\hat{\boldsymbol{z}}\}; \qquad R_1 \leq \rho \leq R_2;\ |z| \leq L. \tag{50}$$

Note that the bound current-density *inside* the disk, i.e., $[\rho^{-1}M_\phi(\rho) + M'_\phi(\rho)]\hat{\boldsymbol{z}}$, would vanish if $M_\phi(\rho) = M_o/\rho$, leaving only radial surface-currents on the disk's top and bottom facets, and also vertical surface-currents on its inner and outer cylindrical walls. In general, the bound-currents of Eq. (50) are subject to the Lorentz force exerted by the total *B*-field, which is the sum of $\mu_o\boldsymbol{H}(\boldsymbol{r})$ produced by $\boldsymbol{J}_{\text{free}}(\boldsymbol{r})$, and $\boldsymbol{B}(\boldsymbol{r}) = \boldsymbol{M}(\boldsymbol{r})$ produced by the magnetization of the disk. (Note that the discontinuity of $\boldsymbol{M}(\boldsymbol{r})$ at the surfaces of the disk introduces a factor of ½ in the corresponding force expressions at these surfaces.) Overall, the self-force $\boldsymbol{J}_{\text{bound}} \times M_\phi(\rho)\hat{\boldsymbol{\phi}}$ gives rise to internal stresses that tend to tear at the fabric of the disk, even though the integral of the self-force over the entire volume of the disk is always equal to zero.

The radial surface-currents on the top and bottom facets of the disk experience equal forces from the azimuthal *H*-field of Eq. (49), yielding

$$\int_{\rho=R_1}^{R_2}\int_{z=-\infty}^{\infty} 2\pi\rho(\boldsymbol{J}_{\text{bound}} \times \mu_o \boldsymbol{H})\,\mathrm{d}\rho\mathrm{d}z = -4\pi L J_o \hat{\boldsymbol{z}} \int_{R_1}^{R_2} M_\phi(\rho)\,\mathrm{d}\rho. \tag{51}$$

The above force is seen to be equal and opposite to that given by Eq.(48). Thus the net Lorentz force exerted on the disk (including the force on the free and bound currents flowing radially and vertically inside the disk) is zero.

Up to this point, we have not considered the *H*-field produced by the wires through which $J_{\text{free}}$ is injected into the disk. The easiest way to handle this problem is to assume that two hollow cylinders of infinite length, one of radius $R_1$, the other $R_2$, carry the input/output current along the *z*-axis. The current-density in both cylinders is an odd function of *z*, moving toward the disk in the small cylinder and away from the disk in the large cylinder. The vector potential produced by the currents of these cylinders is directed along the *z*-axis, is circularly symmetric, and has odd symmetry along *z*. The *H*-field produced by the cylinder currents is thus directed along $\hat{\boldsymbol{\phi}}$ and exhibits the same symmetry as the *H*-field of the radial current within the disk. The conclusions of the preceding paragraphs, therefore, remain intact.

The ponderomotive force of Einstein and Laub is much easier to analyze. The only contributions to this force-density come from the terms $J_{\text{free}} \times \mu_0 H$ and $(M \cdot \nabla) H$, both of which yield a net force of zero on the disk. In fact, $J_{\text{free}} \times \mu_0 H$ integrates to zero along the *z*-axis because of the odd symmetry of the *H*-field with respect to the mid-plane of the disk, while $(M \cdot \nabla) H = -(M_\phi H_\phi / \rho) \hat{\boldsymbol{\rho}}$ integrates to zero because of the odd symmetry of $H_\phi$ along *z* as well as the radial symmetry of the problem.

**9. Concluding remarks**. It has been argued by some that the Lorentz force should be treated as a 4-vector if momentum conservation and the principles of special relativity are to be upheld [11-14]. While this is possible (and helpful) in some instances, it is impracticable in many other situations where alternative inertial frames cannot be identified. Recall that the standard expression of the Lorentz law as given by Eq.(1) specifies a 3-vector force-density in terms of the action of the *E* and *B* fields on charge and current densities, $\rho$ and *J*. Since *E* and *B* form a 2$^{\text{nd}}$ rank tensor that can be transformed from one inertial frame to another, and since $\rho$ and *J* form a 4-vector that can similarly be transformed between inertial frames, one expects the conventional (i.e., 3-vector) Lorentz force law to remain valid in all inertial frames. In other words, Eq.(1) should apply whenever $\rho$, *J*, *E* and *B* are specified, irrespective of the manner in which such specification has come about. Within a given inertial frame, once the sources $\rho_{\text{total}}(\boldsymbol{r}, t)$ and $J_{\text{total}}(\boldsymbol{r}, t)$ are identified, it is generally impossible to write the equations of motion in terms of a force 4-vector, as there is no preferred frame of reference from which one could "import" the force 4-vector in a meaningful and unambiguous way.

As an alternative, authors who are aware of the problems associated with the Lorentz law in the presence of magnetic media, have advocated the use of a hidden energy flux $\mu_0^{-1} M \times E$ and a hidden momentum-density $\varepsilon_0 M \times E$ to circumvent the troubles caused by a straightforward application of the Lorentz law [13,38,41-43]. This approach, if applied consistently, would comply with special relativity and with the conservation laws, but has the disadvantage of introducing hidden entities into theoretical analyses – entities that are inaccessible to direct measurement.

Our approach to this problem has been to eliminate hidden entities from the expressions of EM energy flux and momentum-density, thus restoring the Poynting vector to its standard form, $S = E \times H$, and the EM momentum-density to its Abraham form, $p_{\text{EM}} = E \times H / c^2$. However, rather than relying on the Lorentz force minus the hidden momentum contribution, i.e., Eq.(7), we have advocated the use of the Einstein-Laub force- and torque-density expressions given by Eqs.(8) and (9), respectively [44-50]. This is not entirely unprecedented, as some other authors have also suggested alternative force-density expressions that are either similar or identical to the Einstein-Laub equations [41,42,51-53]. Even in the absence of magnetic materials, the expression most often used in radiation pressure calculations [22-26,28] is $\boldsymbol{F}(\boldsymbol{r},t) = (\boldsymbol{P} \cdot \boldsymbol{\nabla})\boldsymbol{E} + (\partial \boldsymbol{P}/\partial t) \times \mu_o \boldsymbol{H}$, which is readily recognized as the Einstein-Laub formula for a material medium consisting entirely of electric dipoles. What is generally overlooked in the literature is the fact that these alternative expressions are *not* derived from the Lorentz law and that, therefore, some justification is needed if one is to depart from this fundamental law of nature. The goal of the present paper has been to draw attention to the similarities and differences between the Lorentz force law and the Einstein-Laub force- and torque-density equations.

Finally, it must be pointed out that other EM force-density expressions due, for example, to Helmholtz, Minkowski, Abraham, and Peierls [41,54-57], have been proposed and investigated in the past. Reference [54], in particular, contains a wealth of information on Abraham, Minkowski, Einstein-Laub, Helmholtz, and Peierls theories, presenting the relevant theoretical arguments as well as experimental comparisons among the various formulations. The focus of the present paper, however, has been exclusively on the Lorentz and Einstein-Laub formulations, as both these theories are microscopic (as opposed to phenomenological), and can be applied under general circumstances. The Lorentz and Einstein-Laub theories allow polarization and magnetization to be related to electric and magnetic fields in nonlinear, nonlocal, anisotropic, dispersive, and hysteretic materials, without hampering one's ability to predict local force and torque densities. In contrast, the other formulations are restricted to linear media, where $\boldsymbol{P}(\boldsymbol{r},t)$ is proportional to $\boldsymbol{E}(\boldsymbol{r},t)$ and $\boldsymbol{M}(\boldsymbol{r},t)$ is proportional to $\boldsymbol{H}(\boldsymbol{r},t)$. [If these linear media also happen to be free from dispersion, the proportionality constants will be the electric and magnetic susceptibilities $\varepsilon_o \chi_e = \varepsilon_o(\varepsilon-1)$ and $\mu_o \chi_m = \mu_o(\mu-1)$. Here $\varepsilon$ and $\mu$ are the relative permittivity and permeability of the material medium.]

In Minkowski's theory, the force-density only acts where $\boldsymbol{\nabla}\varepsilon$ or $\boldsymbol{\nabla}\mu$ is nonzero (typically at surfaces and interfaces), and there are no forces inside a homogeneous medium. The Abraham force-density is identical to that of Minkowski in stationary situations. However, unlike the Einstein-Laub formulation, electrostriction and magnetostriction appear neither in Abraham's nor in Minkowski's theory, and must be introduced separately, resulting in what is commonly known as the Helmholtz force [54]. In this respect, the Helmholtz force is similar, though by no means identical, to the Einstein-Laub force. The present paper has sought to illustrate, through simple examples, the changes in the force-density distribution that could arise when one "microscopic" force equation (Lorentz) replaces another (Einstein-Laub)–despite the fact that the total force and total torque acting on an isolated object remain the same in the two formulations. In these examples, the electric and magnetic dipoles contained in the material media were permanent (as opposed to induced), thus precluding a description in terms of the $\varepsilon$ and $\mu$ parameters. Consequently, the "phenomenological" formulations of Minkowski, Abraham, and Helmholtz are not general enough to apply to these and many similar situations.